\documentclass[12pt]{article}
\usepackage{graphicx}
\usepackage{amsfonts}
\usepackage{amssymb}
\usepackage{amsmath}
\usepackage{esint}
\usepackage{hyperref}

\def\pbnr{}
\def\speaker{M. Papinutto$^1$,  F. Piccinini$^\dag$, A. Pilloni$^1$, N. Tantalo$^*$,  \underline{A.D. Polosa$^1$}}
\def\onbehalfof{}
\def\title{A Tentative Description of $Z_{c,b}$ States in Terms of \newline Metastable Feshbach Resonances}
\def\affiliation{$^1$Dipartimento di Fisica and INFN,
Sapienza Universit\`a di Roma, \\ Piazzale A. Moro 5,  I-00185 Roma, Italy\\
$^\dag$INFN Sezione di Pavia, Via A. Bassi 6, I-27100 Pavia, Italy\\
$^*$Dipartimento di Fisica and INFN, Universit\`a di Roma ``Tor Vergata'',\\
Via della Ricerca Scientifica 1, I-00133 Roma, Italy 
}
\def\support{The workshop was supported by the University of Manchester, IPPP, STFC, and IOP}

\newcommand\pubnumber{\pbnr}
\newcommand\pubdate{\today}
\def\Title#1{\begin{center} {\Large #1 } \end{center}}
\def\Author#1{\begin{center}{ \sc #1} \end{center}}

\newcommand{\OnBehalf}[1]{\sbox0{#1}\ifdim\wd0=0pt
        {}
	\else
	{\\on behalf of #1}
	\fi}
\newcommand{\SupportedBy}[1]{\sbox0{#1}\ifdim\wd0=0pt
        {}
	\else
	{\footnote{#1}}
	\fi}
\def\Address#1{\begin{center}{ \it #1} \end{center}}

\newcommand\pubblock{\includegraphics[width=5cm]{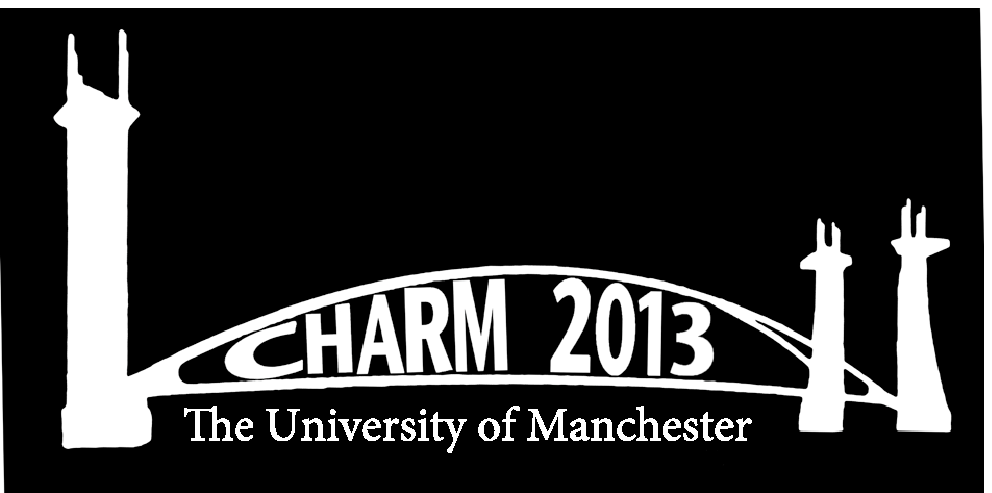}\hfill{\begin{tabular}{l} \pubnumber\\
         \pubdate  \end{tabular}}}
\newenvironment{Abstract}{\begin{quotation}  }{\end{quotation}}
\newenvironment{Presented}{\begin{quotation} \begin{center} 
             PRESENTED AT\end{center}\bigskip 
      \begin{center}\begin{large}}{\end{large}\end{center} \end{quotation}}

\def\venue{The 6$^{th}$ International Workshop on Charm Physics\\
(CHARM 2013)\\
Manchester, UK,  31 August -- 4 September, 2013}

\def\beq{\begin{equation}}
\def\eeq#1{\label{#1}\end{equation}}
\def\eeqn{\end{equation}}

\def\beqa{\begin{eqnarray}}
\def\eeqa#1{\label{#1}\end{eqnarray}}
\def\eeqan{\end{eqnarray}}

\let\bar=\overbar

\def\Dslash{\not{\hbox{\kern-4pt $D$}}}
\def\dslash{\not{\hbox{\kern-2pt $\del$}}}

\def\msb{{\bar{\ssstyle M \kern -1pt S}}}

\begin{document}
\begin{titlepage}
\pubblock

\vfill
\Title{\title}
\vfill
\Author{\speaker\SupportedBy{\support}\OnBehalf{\onbehalfof}}
\Address{\affiliation}
\vfill
\begin{Abstract}
We attempt a description of the recently discovered  $Z_{c,b}$ states  in terms of Feshbach resonances arising from the interaction between the `closed' subspace of hadrocharmonium levels and the `open' one of open-charm/beauty thresholds. We show how the neutrality of the $X(3872)$ might be understood in this scheme and  provide a  preliminary explanation of the pattern of the measured total  widths of $X,Z_{c,b}$.   
\end{Abstract}
\vfill
\begin{Presented}
\venue
\end{Presented}
\vfill
\end{titlepage}
\def\thefootnote{\fnsymbol{footnote}}
\setcounter{footnote}{0}
%

\section{Introduction}
After the most recent findings by BES~\cite{bes} and Belle~\cite{belle}, and after years of experimental research on $X,Z$ exotic spectroscopy, we can rely on a  set of well established states which present very similar features and pose a clear problem about their interpretation. 
The most studied  is the $X(3872)$, a $1^{++}$ state with apparently {\it no} close in mass charged partners.  
The $X(3872)$ happens to be exactly at the $D^0 \bar D^{*0}$ threshold -- a very loosely bound state according to several authors.  

Most recently, very close to the $D^\pm \bar D^{*0}$ threshold, but {\it sensibly above it} by about $20$~MeV, a $1^{+-}$ charged state has been discovered: the $Z_c^\pm(3900)$~\cite{bes,belle}. 
Some hints of an almost degenerate $Z_c^0$ neutral partner seem encouraging~\cite{cleo}.  In spite of the positive binding energy,  the $Z_c$ is described  in several papers as a hadron molecule, a relative of the $X(3872)$~\cite{han2}. 
The issue with  molecules is wether nuclear forces between color singlets are enough binding to explain the large production cross sections at hadron colliders~\cite{bigna}. On the other hand such Van der Waals-like  forces will not be flavor blind and could give rise to SU(3) and even SU(2) incomplete multiplets (as is the case of the $X$). 
Yet, are there alternative explanations of the $X(3872)$ neutrality not relying on the complications of inter-hadron forces?

Sligthy above $Z_c$, close to the $D^*\bar D^*$ threshold, another resonance, the $Z^\prime_c(4025)$ has been found, likely carrying $1^{+-}$ quantum numbers~\cite{bes2}. This state is  {\it above threshold} as well, by about 5~MeV. 
It is still unclear if there are indeed two almost degenerate states  observed in two different decay modes: $D^*\bar D^* $ and $h_c\pi$, with similar widths. However, the latter decay should imply $P=-$, if in $S$-wave.  
We shall assume here that there is only one $Z_c^\prime(1^{+-})$ state, decaying into $h_c \pi$ in $P$-wave. 

Similarly, there are two more $1^{+-}$ states, very close to $B\bar B^*$ and $B^*\bar B^* $ thresholds: $Z_b(10610)$ and $Z_b^\prime(10650)$~\cite{zb}.  These look very much like the partners in the $b$-sector of $Z_c$ and $Z_c^\prime$, but closer to their `related' thresholds. 

\section{An alternative interpretation of the\newline  incomplete $X$ multiplet}
One of the most striking facts to deal with is that $X$ 
has no charged partners, whereas $Z_c$s or $Z_b$s, are observed in three charged states. 
The $Z_c$ particle, its mass and quantum numbers, were predicted with surprising accuracy in the diquark-antidiquark model~\cite{noi} together with a lower partner (which for the moment cannot be excluded). Unfortunately the tetraquark model, in its diquark-antidiquark incarnation, also predicts $G=-$ charged partners of the $X$ and an hyperfine splitting between two neutral $X$ particles almost degenerate in mass. The experimental exclusion of charged partners is rather compelling, but not yet decisive~\cite{maianier}.

In several papers all of these states are described in terms of hadron molecules -- even if their binding energy is positive. Why should such resonances, having the same nature, occur in different charge configurations? Is this due to the (uncontrolled) complexity of nuclear forces binding them or can we formulate alternative explanations? 

Here we consider  the hypothesis that the $X$, as well as $Z$ resonances,  are indeed  compact tetraquark states of the kind $c\bar c q \bar q$, although we will not  stick to the diquark-antidiquark description.  The four quarks are  free to rearrange in two ways: $i)$ a closed charm configuration in which the $J/\psi$, or any heavier charmonium, is surrounded by light quak matter -- what Voloshin calls an {\it hadrocharmonium}~\cite{volo}, $ii)$ an open charm configuration of two barely interacting $D$ mesons~\cite{mols,braaten}. These two alternatives are superimposed and the compact tetraquark (as long as we can consider it as `stable') is described by  
$|\Psi\rangle=|\Psi_P\rangle+|\Psi_Q\rangle$, 
$P$ and $Q$ being the labels of the Hilbert subspaces of open and closed charm states respectively. 

\begin{figure}[t]
\centering
\includegraphics[width=0.4\textwidth]{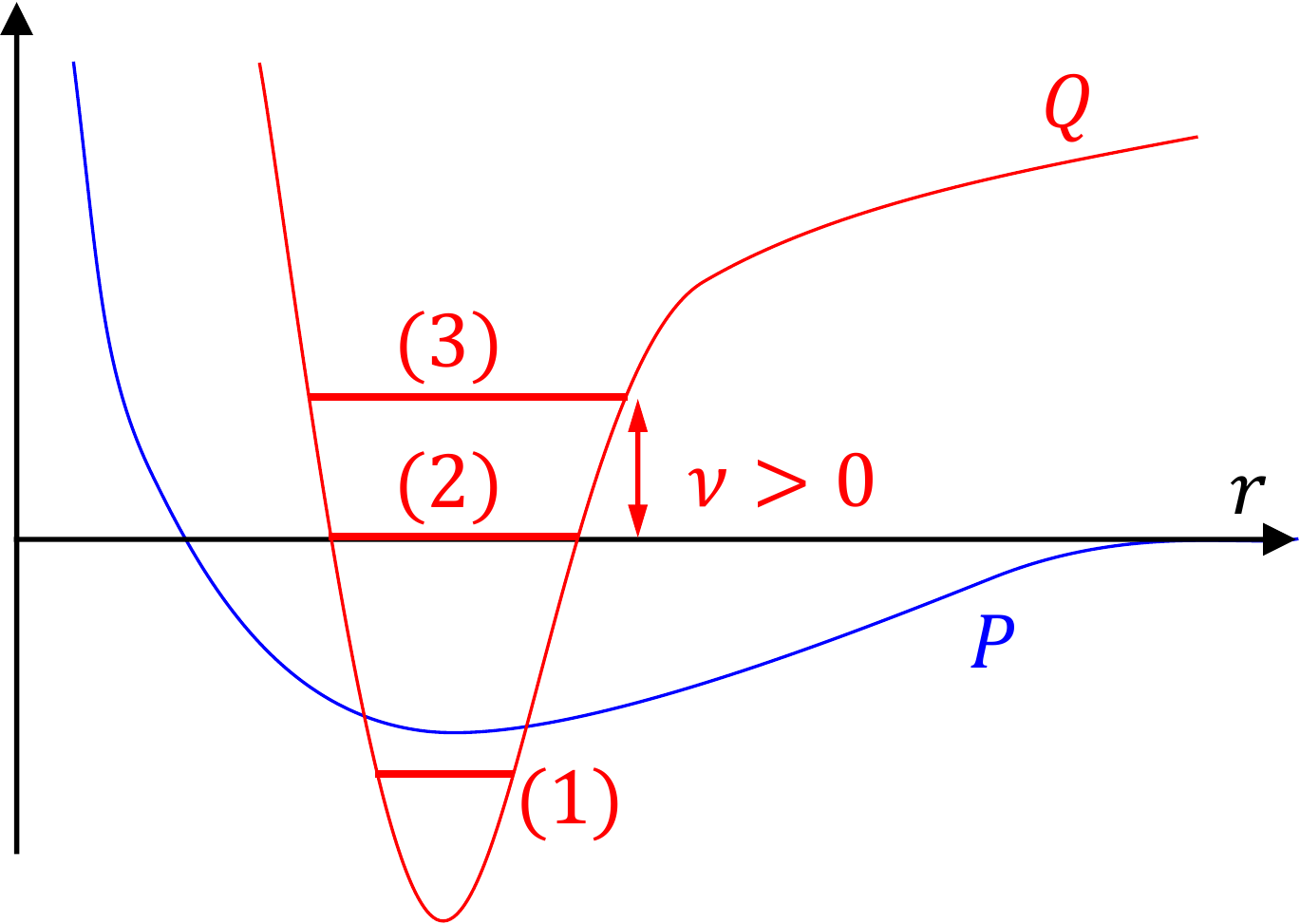}
\caption{\small $P$ and $Q$ are the open (shallow) and closed channels respectively. (1) Charged $X(3872)$ partners suppression, (2) $X$ case (3), $Z_{c,b}$ case.
}
\label{fig:mol}
\end{figure}

We assume that the hadrocharmonium system admits bound states giving rise to a discrete spectrum of levels. A resonance occurs if one of such levels falls close 
to some open-charm threshold. An interaction between these two channels is understood.

The level  we name as `$J/\psi \,\rho$' has $1^{++}$ quantum numbers: it consists of a $c\bar c$ pair with the quantum numbers of the $J/\psi$ and a light component with the quantum numbers of the $\rho$, held together by  hadronic Van der Waals-like forces. The masses of the two components do not have to be {\it a priori} equal to the masses of the `constituent' hadrons, here $J/\psi$ and $\rho$. On the other hand,  note  that $J/\psi \,\rho^0$ would have a mass of 3872~MeV if the masses of the two constituents were just summed. 
If the closest hadrocharmonium level in the $Q$ subspace happens to be above the onset of the continuum spectrum of levels in the $P$ subspace, the coupling between channels gives rise to an attractive interaction and favours the formation of a metastable (Feshbach) resonance  at the hadrocharmonium level. The effect is enhanced the smaller the difference in energy $E-E_{\rm{th}}$, $E$ being the bound state level and $E_{\rm{th}}$ the open-charm threshold energy~\cite{bec}.  
The Feshbach phenomenon is therefore the formation of a resonance in the `scattering' between different internal tetraquark states.

Because of the interaction between channels $P$ and $Q$ we expect  level repulsion and a consequent increase of the detuning parameter $\nu=E-E_\text{th}$, represented in Fig.~1. If the unperturbed detuning between  the $J/\psi \,\rho^\pm$  level and the $D^\pm\bar D^{*}$ threshold is small and negative, as one might guess using naively the central values of the masses of the constituent mesons 
($M(\rho^\pm)-M(\rho^0)\approx 0.7\pm 0.8$~MeV, $M(D^\pm)-M(D^0)\approx 4.8\pm 0.1$~MeV), the interaction between channels will tend to make it more negative. On the other hand, as stated above, this would give rise to a repulsive interaction (the Fermi scattering length $a$ gets a positive correction) likely destabilizing the system and favoring its fall-apart decay. In this case, a bound state solution could be found if the open channel had discrete levels. Indeed such solution would predominantly be in the open channel with a radius $\sim |\nu|^{-1}\approx 40$~fm, much larger than in the original closed channel. We might infer that 
no charged partners of the $X$ have been observed because of these reasons.

The coupling between the $P$ and $Q$ subspaces, described by some $H_{QP}$ Hamiltonian term, appears in the expression of the scattering length 
\begin{equation*}
a\simeq a_P+C\sum_n\frac{|\langle \Psi_n|H_{QP}|\Psi_{{\rm th}}\rangle|^2 }{E_{\rm{th}}-E_n} \simeq a_{NR} - C\frac{|\langle \Psi_{\rm res}|H_{QP}|\Psi_{{\rm th}}\rangle|^2 }{\nu}
\end{equation*}  
where $C$ is some (positive) constant and $a_P$ is the scattering length (at zero energy) when coupling between open and closed channels is neglected. 
Including higher order terms, leading to $H_{QQ}$ interaction terms, the energy of bound states gets shifted (real part of the correction)
and a width is introduced (imaginary part) {\it if} decay into open channel is possible (into levels higher than the threshold one): a metastable state is formed. The last term in the scattering amplitude will then be $\propto 1/(\nu +i\Gamma/2)$. For $\Gamma \ll |\nu|$ the scattering amplitude has a steep $1/\nu$ behavior otherwise, for $\Gamma/2 \gtrsim |\nu|$, the divergence is  smoothed. Consider also that the actual detuning $\nu$ has been shifted from the unperturbed one (being larger in absolute value). 
 
In the case of the $X$,  given that the unperturbed detuning is $\nu\sim 0$ (assuming the mass $J/\psi \,\rho^0$ to be $3872$~MeV), the second order correction to the energy level should be proportional to $\lim_{\epsilon\to 0} \fint_{0}^{2\epsilon} d\nu  |f(\nu)|^2/(\nu-\epsilon)=0$~\footnote{The integral is in the sense of the Principal Value. The function $f(\nu)$ has to be  some regular function in $\nu$. Consider that the less prominent term corresponding to $J/\psi \;\omega$ in the closed channel could also be taken into account although the most simple approximation is that of reabsorb all the non-resonant closed channel states (including the continuum) in $a_{NR}$.}. This should leave the detuning small, $\nu\to 0^+$,
enhancing  the Feshbach resonance phenomenon and making the $X$ a metastable state with very long lifetime~\footnote{A Feshbach mechanism between the open charm spectrum and the $\chi_{c1}(2P)$ charmonium level was first proposed in~\cite{braaten} to explain the narrowness of the $X(3872)$. More precise estimates for the mass of the $\chi_{c1}(2P)$ made this hypothesis unlikely.} -- being essentially driven by the instability of $D^*$, $\Gamma_{D^*}\sim 100$~keV. In this sense the $X$ is a particularly fine-tuned state and we do not expect to have its analog in the beauty sector. 

\section{$Z_c^{(\prime)}$ and $Z_b^{(\prime)}$}
The same line of reasoning could apply to the charged states. The quantum numbers of the $Z_c$ are $1^{+-}$, thus we could choose a hidden charm combination $\psi(3770)\pi^{0,\pm}$ which  falls  about $10$ MeV above the $Z_c$ mass. 
This could be the reason why this resonance, differently from the $X(3872)$, is observed in all  charged states.  Moreover, for the reasons discussed above,  we expect it to be broader than the $X$.  In fact $Z_c$ has a total width of $\approx 50$~MeV to be compared with $\Gamma_X\lesssim 1$~MeV.

As for the $Z_c^\prime$ we might consider a $h_c(2P)\pi$ hadrocharmonium level. The $h_c(2P)$ has not been observed yet but its mass, according to a rough order of magnitude estimate, might be found with the $h_c(1P)+h_b(2P)-h_b(1P)$ mass formula (locating $h_c(2P)\pi$ at 4025~MeV, just a few MeVs above the threshold value $\bar D^* D^{*\pm}$ at 4017~MeV).
As a {\it caveat} we notice that the $h_c(2P)\pi$ hadrocharmonium level would have negative parity if no unit of orbital angular momentum is introduced. The decay $Z_c^\prime \to h_c(1P)\pi$, if confirmed to be $P$-wave, would support this hypothesis~\footnote{On the other hand, orbital angular momentum could have the effect of rising the mass of the state.}. 

As we do not yet know the hadrocharmonium spectrum, one might wonder how the scheme here described might be preliminarily tested. As a first step we  apply the Feshbach formalism to estimate the width of the states. 
According to the Fermi golden rule, the width of a metastable Feshbach resonance, with $\nu>0$, has to be proportional to the density of accessible levels $dn/dE$ thus leading to $\Gamma\simeq A\sqrt{\nu}$: 
when $\nu\to 0^+$ the two channels will be strongly hybridized and the resulting metastable state be very narrow~\footnote{The width formula is, for $\nu>0$, given by $\Gamma(\nu)\sim \sqrt{\nu_c \nu}$  where the critical value of the detuning $\nu_c$ is a function of the details of the potentials involved, including the hybridizing one.  Narrow width approximation would require $\Gamma(\nu)\lesssim\nu$. We have to relax it to $\Gamma(\nu)<C\nu$ where $C$ is some positive constant; in other words we  need $\nu>\nu_c/C^2$. In our computation (see the text) we estimate a critical value $\nu_c\sim 100$~MeV, thus $C\sim 2$ is sufficient when considering the $Z_c^{(\prime)}$, but we would need $C\sim 10$ for the $Z_b^{(\prime)}$.
If $\nu<0$ and the open channel admits discrete levels, one can indeed find a bound state in the open channel. 
We assume to be in the conditions where the open channel has no discrete levels, {\it i.e.} the pion forces are not enough binding~\cite{rux}.
}. 

Hence, the detuning for $Z_c$ is $\nu \approx 20$~MeV and $\Gamma_{\rm exp}\simeq 46\pm 22 $~MeV. This would lead to $A\simeq 10\pm 5~{\rm MeV}^{1/2}$. 
As for the  $Z_c^\prime$, the detuning is roughly estimated to be $\nu \approx 9$~MeV, which would then lead to  $\Gamma\approx 29\pm 16$~MeV, using $A$ computed above. 
The measured width is $\Gamma_{\rm exp}\simeq 24.8\pm 9.5$~MeV (at a mass of 4026~MeV in the decay channel $D^*\bar D^*$), sensibly smaller than the measured width of $Z_c$ and in good agreement with our simple estimate. This is also compatible with the $Z_c^\prime \to h_c \pi$ channel, having a mass of 4023~MeV but a width of $\Gamma_{\rm exp}\simeq 7.9\pm 3.7$~MeV. 

Under the hypothesis that the quantum numbers of  $Z_b$ and $Z_b^\prime$ are $1^{+-}$, we might conjecture that the two closed charm states are $\chi_{b0}(1P)\rho$ and $\chi_{b1}(1P)\rho$, with a unit of orbital angular momentum to give the overall positive parity. Both combinations would be only slightly higher than the related $B \bar B^*$ and $B^* \bar B^*$ thresholds (the effect of the orbital motion  on the mass of these hadrobottomonium levels should be estimated). 
In the case of $Z_b$ and $Z_b^\prime$ the detunings are  roughly $\nu=2.7$~MeV and $\nu^\prime=1.8$~MeV respectively.  The measured decay widths reported by Belle are $\Gamma_{\rm exp}=18.4\pm 2.4$~MeV and $\Gamma^\prime_{\rm exp}=11.5\pm 2.2$~MeV for the lighter state $Z_b(10610)$ and the heavier $Z_b^\prime(10650)$ respectively, {\it i.e.} $(\Gamma/\Gamma^\prime)_{\rm exp}\approx 1.6\pm 0.4$, to be compared with $\Gamma/\Gamma^\prime=\sqrt{\nu/\nu^\prime}=1.2$. 
We observe that in the beauty sector $A\approx 10\pm 3~{\rm MeV}^{1/2}$, as for the charm.

More states, together with their widths, might  be predicted in the charm and beauty sector along the same lines. There is however an important selection rule: 
the lifetime $\tau_D$ of the open charm molecular constituents must be confronted with the characteristic time $\tau_F$ of the Feshbach resonance. If $\tau_D \gg \tau_F$ our formulation of Feschbach mechanism can neglect the instability of the constituents. If \mbox{$\tau_D \sim \tau_F$} the decay of the constituents might happen during the lifetime of the molecule, thus challenging the Feshbach mechanism. We could consider only the stablest open flavor mesons $D$,$B$ and $D^*$,$B^*$, reproducing the observed $X$ and $Z_{c,b}^{\left(\prime\right)}$. If we weaken this requirement, for example in charm sector with an $S$-wave hadrocharmonium, the first state we predict appears at the thresholds $D_1^0 \bar D^0$ and $h_c \,\omega$, thus having quantum numbers $\left(I^G\right)\,J^{PC}=\left(0^+\right)\,1^{-+}$, a mass of $\approx 4310$~MeV and a width of $\approx 40$~MeV.

\section{Conclusions}
We confront with the problem of the absence of charged partners of  $X(3872)$ 
and with the recent findings on  the $Z_c$s  and $Z_b$s resonances. We attempt a  description of all these states as Feshbach resonances arising in the scattering between internal states of a  tetraquark configuration. We distinguish between a closed subspace of hadrocharmonium levels and an open subspace of open-charm thresholds. There could be only a few hadrocharmonium-like states almost matching open charm thresholds. The most perfect match is that realized by the $D^0 \bar D^{*0}$ threshold with $J/\psi \,\rho^0$, which we refer to as the neutral $X$. 
The $J/\psi \,\rho^\pm$ might fail to match   $D^\pm \bar D^{*0}$, being below it, and a metastable state might be forbidden. Few more matchings can be found corresponding to $Z_c$s and $Z_b$s. We relate a parameter of the Feshbach resonances, the detuning, to the observed width of the states, reporting some interesting agreement with data.

\section*{Acknowledgements} ADP whish to thank Ben Grinstein for his hospitality at UCSD and for intersting discussions.

\end{document}